\documentclass[aps,prl,10pt,superscriptaddress,showpacs,amsmath,twocolumn,floatfix]{revtex4-1}
\usepackage[dvipsnames]{xcolor}
\usepackage[colorlinks=true,bookmarks=false,linkcolor=NavyBlue,urlcolor=NavyBlue,citecolor=NavyBlue,breaklinks]{hyperref}
\usepackage{amsmath}
\usepackage{amsfonts}
\usepackage{tabularx}
\usepackage{graphicx}
\usepackage{times}
\usepackage{mathtools}
\usepackage{braket}

\newcommand{\real}{\operatorname{Re}}

\newcommand{\parti}[2]{\frac{\partial #1}{\partial #2}}

\newcommand{\intall}{\int_{-\infty}^{\infty}}

\newcommand{\bk}[1]{\left(#1\right)}

\begin{document}

\title{Comment on ``Resurgence of Rayleigh's curse in the presence of partial coherence''}

\author{Mankei Tsang}
\email{mankei@nus.edu.sg}
\homepage{http://mankei.tsang.googlepages.com/}
\affiliation{Department of Electrical and Computer Engineering,
  National University of Singapore, 4 Engineering Drive 3, Singapore
  117583}

\affiliation{Department of Physics, National University of Singapore,
  2 Science Drive 3, Singapore 117551}

\author{Ranjith Nair}
\affiliation{Department of Electrical and Computer Engineering,
  National University of Singapore, 4 Engineering Drive 3, Singapore
  117583}

\date{\today}


\begin{abstract}
  Larson and Saleh [Optica \textbf{5}, 1382 (2018)] suggest that
  Rayleigh's curse can recur and become unavoidable if the two
  sources are partially coherent.  Here we show that their
  calculations and assertions have fundamental problems, and
  spatial-mode demultiplexing (SPADE) can overcome Rayleigh's curse
  even for partially coherent sources.
\end{abstract}

\maketitle
In Ref.~\cite{larson18}, Larson and Saleh suggest that Rayleigh's
curse---as originally defined in Ref.~\cite{tnl}---can recur and
become unavoidable if the two sources are partially coherent. Here we
show that their calculations and assertions have fundamental problems.
First we show that the Fisher information of the
spatial-mode-demultiplexing (SPADE) measurement \cite{tnl} can
overcome Rayleigh's curse as long as the correlation between the two
sources is not too positive, contrary to the claim in
Ref.~\cite{larson18}.  For simplicity, we use the semiclassical theory
described in Refs.~\cite{tnl2,tsang_semiclassical}, which is
equivalent to the quantum formalism for weak thermal optical sources.
For two partially coherent sources, the mutual coherence on the image
plane is
\begin{align}
\Gamma(x,x') &= N_0
\left[h_+(x)h_+^*(x') + h_-(x)h_-^*(x') + 
\right.
\nonumber\\&\quad
\left.\gamma h_+(x)h_-^*(x')+\gamma^* h_-(x)h_+^*(x')\right],
\label{mutual}
\end{align}
where $N_0$ is the expected photon number from one source, $\gamma$ is
the complex degree of coherence, $h_\pm(x) = h(x\pm s/2)$ is the
wavefunction due to each source, and $s$ is the separation
\cite{mandel}. Assume
$h(x) = (\sqrt{2\pi}\sigma)^{-1/2}\exp[-x^2/(4\sigma^2)]$, and
consider the average photon number in a Hermite-Gaussian mode
$\phi_q(x)$ given by
\begin{align}
n_q &= \intall dx \intall dx' \phi_q^*(x)\Gamma(x,x')\phi_q(x').
\end{align}
For the first-order mode with $\phi_1(x) = (x/\sigma) h(x)$,
\begin{align}
n_1 &= 2N_0\bk{1-\real\gamma}\frac{s^2}{16\sigma^2}
\exp\bk{-\frac{s^2}{16\sigma^2}}.
\end{align}
Assuming Poisson statistics, which is the standard assumption for
thermal sources at optical frequencies \cite{goodman_stat,pawley}, the
Fisher information is
\begin{align}
F_q &= \frac{1}{n_q}\bk{\parti{n_q}{s}}^2,
\\
F_1 &= \frac{N_0}{2\sigma^2}\bk{1-\real\gamma} 
\bk{1-\frac{s^2}{16\sigma^2}}^2\exp\bk{-\frac{s^2}{16\sigma^2}}.
\end{align}
Notice that
\begin{align}
  F_1(s = 0) &= \frac{N_0}{2\sigma^2}\bk{1-\real\gamma},
\end{align}
which is zero \emph{only} when $\gamma = 1$, viz., when the two
sources are positively and perfectly correlated.  For all other values
of $\gamma$, $F_1(s=0)$ is positive and Rayleigh's curse is averted.

\begin{figure}[htbp!]
\centerline{\includegraphics[width=0.48\textwidth]{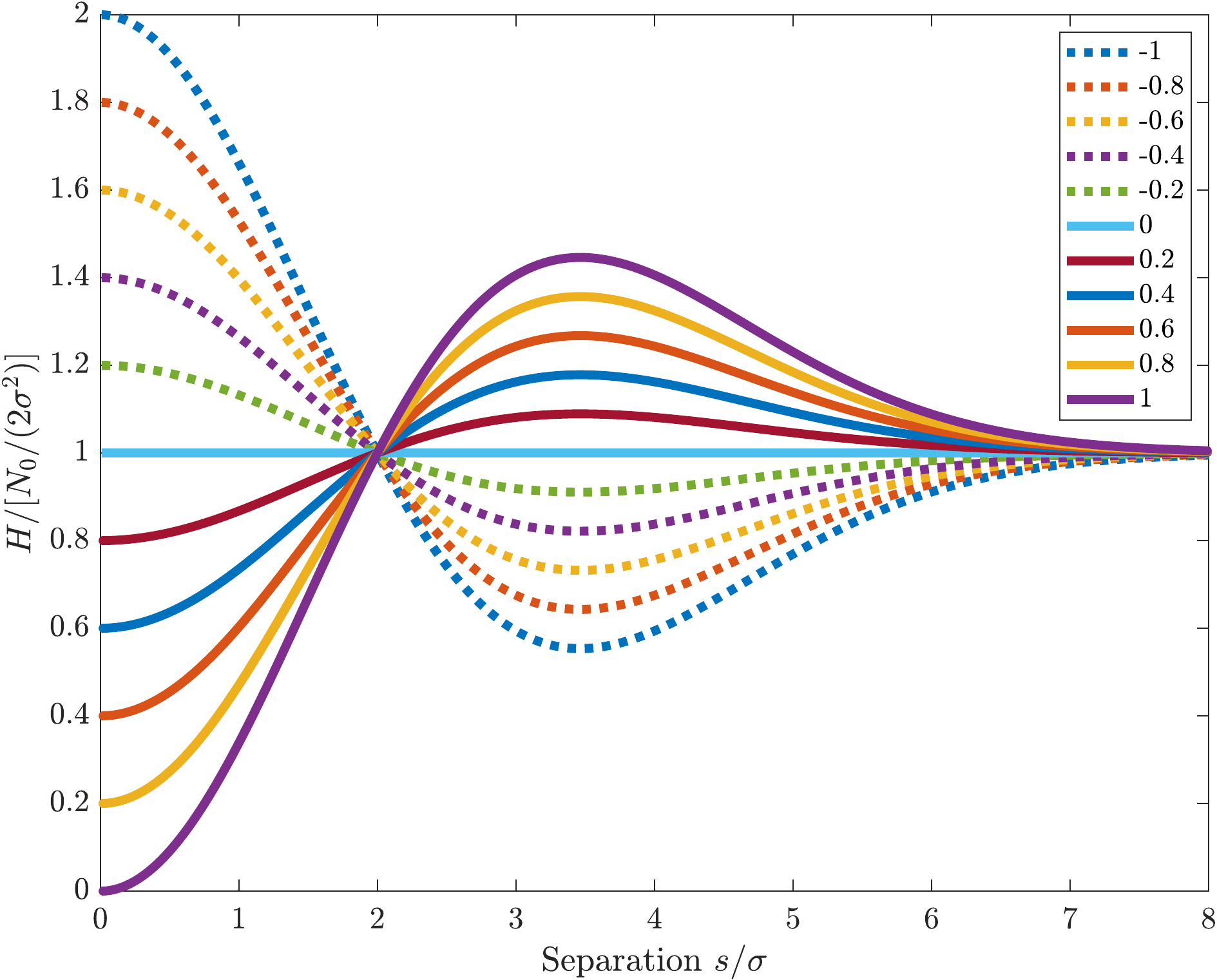}}
\caption{Total Fisher information in Hermite-Gaussian modes up to
  $q = 20$ versus the separation for various values of $\gamma$
  (denoted by the legend).}
\label{partial_spade}
\end{figure}

The total Fisher information $F = \sum_{q=0}^\infty F_q$ achievable by
SPADE must be even higher. Figure~\ref{partial_spade} plots the Fisher
information---numerically computed by summing up the information in
modes up to $q = 20$---for various values of $\gamma$.  The curves
vary smoothly for varying $\gamma$ and possess a pleasing symmetry.
For $\gamma = \pm 1$, the curves are also consistent with
Ref.~\cite{localization} in the context of coherent sources.  Most
importantly, for anticorrelated sources ($\gamma < 0$), the
information does not vanish for small $s$ and does not suffer from
Rayleigh's curse, unlike the behavior suggested by Fig.~3 in
Ref.~\cite{larson18}.

In fact, Fig.~\ref{partial_spade} even shows an enhancement for
sub-Rayleigh anticorrelated sources.  This is consistent with the
intuitive explanation of how SPADE enhances the Fisher information
\cite{tnl}: For $s \ll \sigma$, the $q = 1$ mode is the most sensitive
to the separation parameter, while the $q = 0$ mode, which contributes
mostly background noise to direct imaging, is filtered by SPADE. If
the sources are close and anticorrelated, the coupling to the $q = 1$
mode is enhanced, so the information is also enhanced.

\begin{figure}[htbp!]
\centerline{\includegraphics[width=0.48\textwidth]{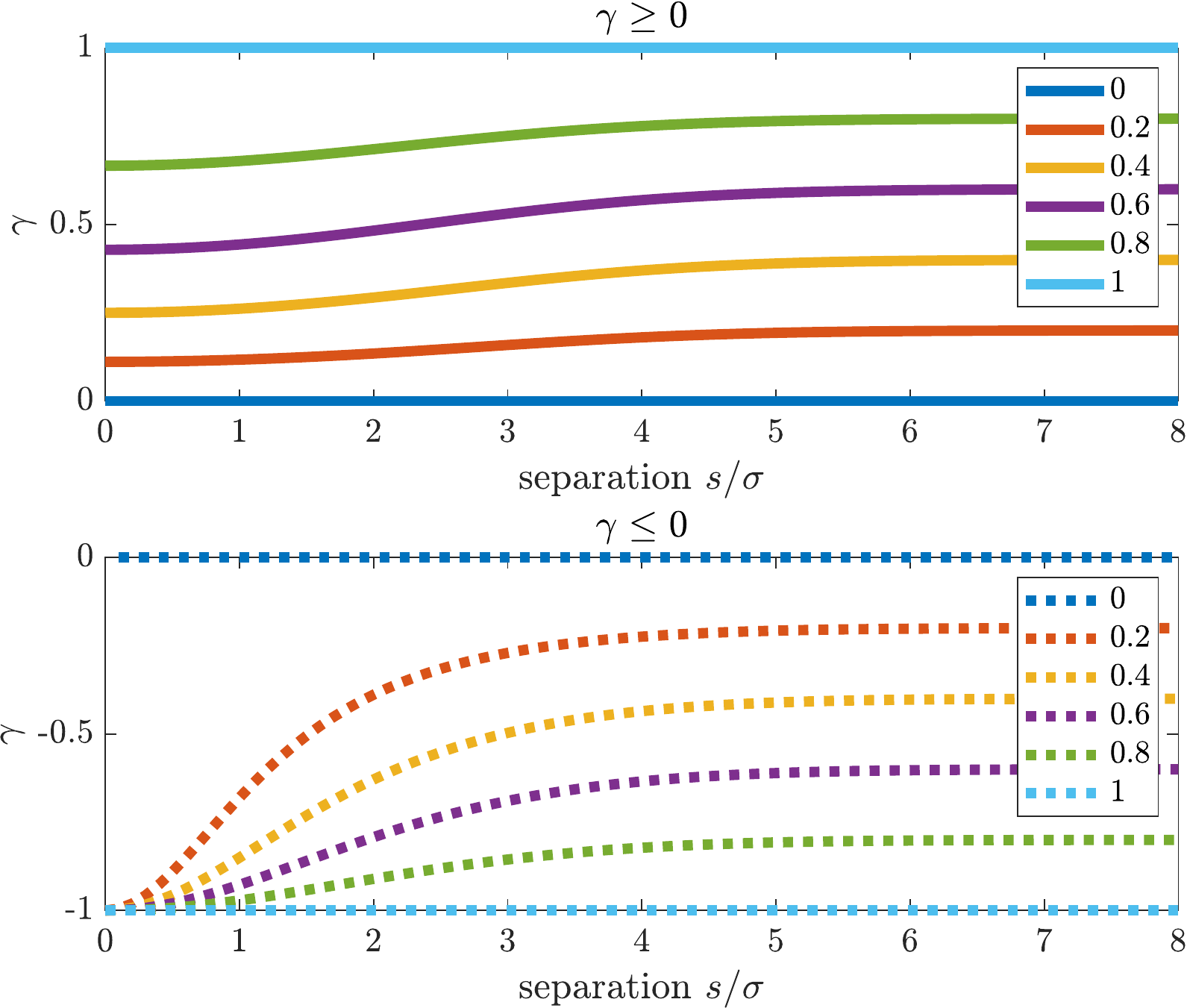}}
\caption{Plots of the degree of coherence $\gamma$ versus separation
  $s$ for various values of $p$ (denoted by the legends), according to
  Eq.~(13) in Ref.~\cite{larson18}. The Gaussian point-spread function
  is assumed.}
\label{larson}
\end{figure}

There are at least two problems with Ref.~\cite{larson18} that can
explain the disagreement.  One is parametrization. Instead of dealing
with the degree of coherence $\gamma$ directly, Ref.~\cite{larson18}
defines another parameter $p$, which is related to $\gamma$ through
Eq.~(13) in Ref.~\cite{larson18}. Every curve plotted in
Ref.~\cite{larson18} assumes a fixed $p$ rather than $\gamma$. The
problem is that fixing $p$ leads to an unphysical dependence of
$\gamma$ on the separation, as shown in Fig.~\ref{larson}. It is
unclear what sources in practice can exhibit such behaviors, which
look especially bizarre in the case of $\gamma < 0$: the sources
acquire significant anticorrelation as they get closer and achieve
perfect anticorrelation at $s = 0$, even for moderate levels of
$p$. Thus any result that assumes a fixed $p$ can be misleading.

The other problem with Ref.~\cite{larson18} is its use of a normalized
one-photon density operator in the computation of the quantum Fisher
information (QFI). While the use is justified for incoherent sources
\cite{tnl,ant,*tsang16c,*tsang18}, it can lead to incorrect results
otherwise. For a weak thermal state with $M$ temporal modes, the
density operator for each temporal mode can be approximated as $\rho =
(1-\epsilon)\rho_0 + \epsilon\rho_1$, where $\rho_0$ is the vacuum
state, $\rho_1$ is the one-photon state, and $\epsilon\ll 1$ is the
expected photon number per temporal mode, given by $\epsilon = \intall
dx \Gamma(x,x)/M$ \cite{tnl}.  For the incoherent sources assumed in
Refs.~\cite{tnl,ant,*tsang16c,*tsang18}, $\epsilon$ does not depend on
the parameters of interest $\theta$, so the QFI in $\rho$, defined as
$Q(\rho)$, is simply $\epsilon Q(\rho_1)$.  For partially coherent
sources, however, $\epsilon$ can depend on the parameters because of
interference.
With $\rho_0$ being independent of any parameter and $\rho_0$ and
$\rho_1$ living in orthogonal subspaces, it is not difficult to show
that
\begin{align}
Q(\rho) &= \epsilon Q(\rho_1) + J(\epsilon),
\end{align}
where $J$ is the classical information for the distribution
$\{1-\epsilon,\epsilon\}$ given by
\begin{align}
J_{\mu\nu}(\epsilon) &= \frac{1}{\epsilon(1-\epsilon)}
\parti{\epsilon}{\theta_\mu}\parti{\epsilon}{\theta_\nu}
\approx
\frac{1}{\epsilon}
\parti{\epsilon}{\theta_\mu}\parti{\epsilon}{\theta_\nu}.
\end{align}
Even if the QFI is evaluated on a per-photon basis as
$Q(\rho)/\epsilon = Q(\rho_1) + J(\epsilon)/\epsilon$,
$J(\epsilon)/\epsilon$ may not be negligible.  By ignoring
$J(\epsilon)/\epsilon$ and considering only $Q(\rho_1)$,
Ref.~\cite{larson18} must have underestimated the total information
for partially coherent sources.

Our final issue with Ref.~\cite{larson18} is its claim that the
wavelength-scale coherence length of Lambertian sources \cite{mandel}
can be important, when the opposite is much more likely for
fluorescence microscopy and observational astronomy---two of the
biggest applications of incoherent imaging. First of all, the
Lambertian model is well known to be heuristic and does not take into
account the detailed physics of the emitters. It would be a major
surprise if the fluorescent emissions of different particles in common
microscopy could exhibit any cooperative effect and acquire coherence
at the object plane, contrary to the incoherence assumption widely
adopted in fluorescence microscopy \cite{pawley}.  Second, a
wavelength-scale coherence length can hardly be relevant to
observational astronomy, as the numerical aperture ($\textrm{NA}$) is
extremely low and the wavelength $\lambda$ is smaller than the
characteristic length scale $\sigma \sim \lambda/\textrm{NA}$ by many
orders of magnitude. Third, while it is true that spatial coherence
develops in the field during diffraction even for spatially incoherent
sources by virtue of the Van Cittert-Zernike theorem \cite{mandel},
the effect has already been properly incorporated in the model used in
Refs.~\cite{tnl,tnl2,tsang_semiclassical,ant,*tsang16c,*tsang18}, and
one should not confuse this effect with partial coherence \emph{at}
the sources.

In conclusion, spatial coherence of the sources is unlikely to be
significant in key applications of incoherent imaging, and even if it
is, Ref.~\cite{larson18} has overblown its detrimental effect.

This work is supported by the Singapore Ministry of Education Academic
Research Fund Tier 1 Project R-263-000-C06-112.

\bibliography{research}

\end{document}